\documentstyle[floats,aps]{revtex}

\begin{document}
\draft
\catcode`\@=11
\catcode`\@=12
\twocolumn[\hsize\textwidth\columnwidth\hsize\csname%
@twocolumnfalse\endcsname
\title{A Systematic Improvement for Calculation to Conductivity in Anomalous 
Propagation of Surface Acoustic Wave at $\nu=\frac{1}{2}$}
\author{Yue Yu}
\address{Institute of Theoretical Physics, Academia Sinica, Beijing 100080,
P. R. China}

\maketitle
\begin{abstract}

We report a systematic improvement to calculate the conductivity which 
associates to the anomaly of the propagation of surface acoustic waves 
at $\nu=\frac{1}{2}$ above a two-dimensional electron gas. 
We try to resolve the discrepancy between theoretical 
and experimental values for the magnitude of $\sigma_{xx}(q)$ by considering 
the contribution to the response functions from the self-interaction among  
the Chern-Simons gauge fluctuations.
\end{abstract}


\pacs{PACS numbers:71.10.Pm, 73.40.Hm, 73.20.Dx}]

The anomaly of the surface acoustic wave (SAW) propagation above a 
two-dimensional electron gas (2DEG) in GaAs/AlGaAs heterostructures 
\cite{will} provided the evidence for the  existence of the Fermi 
surface of the composite fermions (CF) \cite{Jain} in the Hall metallic 
states at the filling factor $\nu=\frac{1}{2}$ \cite{HLR}. Measurements of   
both absorption and velocity shift of the SAWs showed anomalous maximum and
minimum at $\nu=\frac{1}{2}$ with respect to the prediction of the $dc$ 
theory \cite{will2}. Correspondingly, a maximum of the conductivity exhibits  
at $\nu=\frac{1}{2}$. In the clear range where the wavelength is smaller
than the CF's mean free path, the conductivity is of the linear-dependence
on the wavevector. These observations agree qualitatively with the 
theoretical results in the fermionic Chern-Simons theory to $\nu=\frac{1}{2}$ 
\cite{HLR}. There is  also some progress in the study of the 
agreement between theory and experiment \cite{Oppen}. However, from
the beginning, there has existed a systematic discrepancy between the 
theoretical and experimental values for the magnitude of the conductivity 
$\sigma_{xx}(q)$. In the fermionic Chern-Simons theory, there is no 
adjustable parameter to enhance
the absolute value of $\sigma_{xx}(q)$ which is smaller than that obtained 
in experiments \cite{HLR}.

On the other hand, there has been a series of developments in the theory of the 
Hall metallic states. Read has interpreted the CF as physical vortices  bound
to an electron \cite{Read}. Field theoretically, this bound state notion
could be realized in two approaches. A non-hermitian transformation leads
to the theory has a better saddle point \cite{RR} and a corresponding 
perturbative theory has been developed by Wu and the present author 
\cite{WY}. More recently, 
this better saddle point of the theory has been resolved by using a common 
field theoretical approach \cite{SM}. It is found that the effective mass
for CF is finite in the Hartree-Fock approximation \cite{SM,YSD}. We benefit 
from  use of the temporal gauge to the Chern-Simons gauge fluctuations 
to abstract this meaningful CF effective mass \cite{YSD}. We anticipate
more physical results could be revealed by using this field theoretical 
approach.

In this Letter, we would like to report a systematic improvement to the 
discrepancy mentioned above if we take account of the self-interaction 
among the Chern-Simons gauge fluctuations which are induced by the density 
fluctuations of CF. This self-interaction has been ignored in the work of
Halperin, Lee and Read (HLR) in the random phase approximation (RPA). We will
see that there is a comparable contribution to the response function
from the bubble diagram of the Chern-Simons gauge fluctuation (Fig.1).
We find that this gives rise to an opposite correction to the free CF
current-current response function and so $\sigma_{xx}(q)$ becomes adjustable. 
We calculate this correction and find that it could resolve the discrepancy 
in the conductivity if we use the experimental data in the SAW propagation
\cite{will,will2,will3,will4,will5}. We also suggest a possible measurement 
for the CF effective mass via the SAW propagation.

We would like to study a 2DEG which is placed in a perpendicular 
magnetic field $B$  and embedded in a uniform positive background. 
The electrons' spin is polarized by the strong magnetic field. 
For the two-body interaction potential $V$, the $N$-electron Hamiltonian 
reads
\begin{equation}
H_e=\frac{1}{2m_b}\sum_i\biggl[-i\hbar\nabla_i-\frac{e}{c}\vec{A}_i(\vec{x}_i)
\biggr]^2+\sum_{i<j} V(\vec{x}_i-\vec{x}_j),
\end{equation}
where the vector potential $\vec{A}$ is corresponding to the magnetic field
$B$ and $m_b$ is the band mass of the electrons. Hereafter, we will use the 
unit $\frac{e}{c}=\hbar=1$ except it is explicitly restored. 
Here we do not confine the electrons in the LLL. The attraction between 
the electrons and the uniform background is not
explicitly shown up. 

The fermionic Chern-Simons field theory \cite{FL,HLR,KZ} is based on an
anyon transformation. The mean-field theory leads to the prominent fractional
Hall plateaus for the odd denominator fraction and the Fermi surface for the CF
to $\nu=\frac{1}{2}$. In the second quantization version
for a CF field $\psi_{cf}$ with the gauge fluctuations
around the mean-field state, the corresponding Lagrangian reads
\begin{eqnarray}
L_{cf}&=&\int d^2x\biggl[ \psi^\dagger_{cf}(\vec{x},t)i\partial_t\psi_{cf}
(\vec{x},t)\nonumber\\
&+&\frac{e}{2\pi\tilde\phi}a_i(\vec{x},t)\epsilon_{ij}\partial_t
a_j(\vec{x},t)\biggr]-H_{cf},
\end{eqnarray}
where we have fixed the Chern-Simons gauge fluctuations to the temporal
gauge $a_0=0$ \cite{YSD} and the spatial components obey a gauge invariant
constraint
\begin{equation}
(\nabla\times \vec{a}-2\pi\tilde\phi\delta\rho)|phys>=0.\label{cst}
\end{equation}
Here the even number $\tilde\phi =2$ for $\nu=\frac{1}{2}$.
Recall the transverse component of the Chern-Simons gauge field 
is canonically conjugate to the longitudinal component, one has
\begin{equation}
[a_i,a_j]=\epsilon_{ij}.
\end{equation}
The Hamiltonian in the temporal gauge then reads
\begin{eqnarray}
H_{cf}&=&\int d^2x |(-i\nabla+\vec{a}(\vec{x}))
\psi_{cf}(\vec{x})|^2\nonumber\\
&+&\frac{1}{2}\int d^2x\int d^2x'\delta \rho(\vec{x})
\frac{e^2}{\varepsilon|\vec{x}-\vec{x}'|}\delta\rho(\vec{x}'),
\label{cfh}
\end{eqnarray}
where we have specified the interaction to be the Coulomb interaction.
This Hamiltonian could be decomposed into the following several terms,
in terms of Shankar and Murthy \cite{SM},
\begin{equation}
H_{cf}=H_{0f}+H_{0a}+H_{i}+H_{ia},
\end{equation}
where
\begin{eqnarray}
H_{0f}&=&\frac{1}{2m_b}\int d^2 x |\nabla\psi|^2, \nonumber \\
H_{0a}&=&\frac{n_e}{2m_b}\int d^2 x(a_x^2+a_y^2)\nonumber\\
&+&\frac{1}{8\pi^2
\tilde\phi^2}\int d^2xd^2x'
\nabla\times\vec{a}(\vec{x})V(\vec{x}-\vec{x}')\nabla'\times\vec{a}
(\vec{x}'), \nonumber\\
H_{i}~&=&\int d^2 x \vec{a}\cdot \frac{i}{2m_b}(\psi^\dagger\nabla
\psi-\nabla\psi^\dagger \psi),\nonumber\\
H_{ia}&=&\frac{1}{2m_b}\int d^2x \delta\rho\vec{ a}^2.
\label{hdc}
\end{eqnarray}
Here $H_{0f}$ and $H_{0a}$ stand for the free Hamiltonians of the CF and 
the gauge fluctuations respectively. $H_i$ is the interaction between
the CF and the gauge field while $H_{ia}$ can be rewritten as
\begin{equation}
H_{ia}=\frac{1}{4\pi\tilde\phi m_b}\int d^2x (\nabla\times \vec{a})
\vec{a}^2,
\end{equation}
where we have used the constraint (\ref{cst}) to solve $\delta\rho$. This
term, now, describes the self-interaction among the gauge fluctuations.
The perturbation expansion can be carried out according to the Feynman rules
set up in ref.\cite{YSD}. Here we address the gauge field self-interaction
(Fig.2) as
\begin{eqnarray}                         
&&f_{122}(-\vec{q}-\vec{q},\vec{q}, \vec{q}') =\frac{i}{8\pi m_b}(q_2+q_2'),
\nonumber\\
&&f_{211}(-\vec{q}-\vec{q},\vec{q}, \vec{q}') =\frac{-i}{8\pi m_b}(q_1+q_1').
\end{eqnarray}
In our notation, the suffices `1' and `2' denote the longitudinal and transverse
directions to the wavevector. It has been pointed out that this interaction  
vertex is not renormalizable because it is not relevant to the CF kinetic
energy \cite{SM,YSD}.

In a previous work \cite{YSD}, we have shown that the CF self-energy in the
Hartree-Fock approximation gives a finite effective mass for CF which is
consistent with Shankar and Murthy's result \cite{SM} and encouraged by 
the numerical analysis \cite{MdA},
\begin{equation}
\frac{1}{m^*}=\frac{Ae^2\l_{1/2}}{\varepsilon},\label{em}
\end{equation}
where $A=\frac{1}{6}$ and $l_{1/2}=k_F^{-1}$ is the magnetic length for $\nu=\frac{1}
{2}$. We also calculated the gauge propagator in the RPA and its long
wavelength limit in $\omega\ll v_F q$ is given by
\begin{eqnarray}
&&D_{11}(\vec{q},\omega)=-\frac{2\pi}{m^*}\frac{q^2}{\omega^2-i\delta},\nonumber\\
&&D_{22}(\vec{q},\omega)=-\frac{q}{i\omega\gamma-q^2\chi}, \label{prop}
\end{eqnarray}
where $\gamma=\frac{k_F}{2\pi}$ and $\chi=\frac{e^2}{8\pi\varepsilon}$. 

Before giving our main result, we review briefly several
important facts haven been known \cite{HLR,will}. The experiments of 
the SAW's propagation were performed in 
high quality GaAs/AlGaAs heterostructures while $q\ll k_F$ and 
$\omega\ll v^*_Fq$ ( $v^*_F=k_F/m^*$). At exact $\nu=\frac{1}{2}$,
the real and image parts of
the density-response function $K_{00}(q,\omega)$ in $\omega=v_s q$ give 
rise to the a velocity shift $\Delta v_s$ and an attenuation rate $\kappa$
for the SAW amplitude,
\begin{eqnarray}
&&\frac{\Delta v_s}{v_s}=\frac{\alpha^2}{2}\frac{1}{1+[\sigma_{xx}(q)/
\sigma_m]^2},\nonumber\\
&&\kappa=\frac{q\alpha^2}{2}\frac{[\sigma_{xx}(q)/\sigma_m]}{
1+[\sigma_{xx}(q)/\sigma_m]^2},
\end{eqnarray}
where $\alpha$ is a constant  proportional to the piezoelectric coupling
of GaAs. The constant $\sigma_m=\frac{v_s \varepsilon}{2\pi}$. However,
to compare with the experimental data, it is necessary to use a value
of $\sigma_m$ approximately four times larger than this theoretical one. 
This discrepancy keeps no explained at present. The longitudinal conductivity
has been calculated in \cite{HLR} as
\begin{equation}
\sigma_{xx}(q)=\frac{\rho_{yy}(q)}{\rho_{xy}^2},
\end{equation}
where $\rho_{xy}=4\pi\hbar/e^2$, and
\begin{equation}
\frac{1}{\rho_{yy}(q)}=-e^2\lim_{\omega=v_sq\to 0} \frac{1}{\omega}
{\rm Im}\tilde K_{22}(q,\omega), \label{rhoyy}
\end{equation}
where $\tilde K_{22}(q,\omega)$ includes the diagrams which are irreducible
with respect to the Chern-Simons interaction. The RPA consists of replacing
$\tilde K$ by the the free CF response $K^0$ (e. g., ${\rm Im}K_{22}^0
=-\frac{2n_e \omega}{k_F q}$). In this approximation, 
Halperin et al arrived at \cite{HLR}
\begin{eqnarray}
&\displaystyle\rho_{yy}(q)=\frac{2\pi}{k_F}q\frac{\hbar}{e^2},
& {\rm for}~~~q\gg \frac{2}{l}, \label{clear}\\
& \displaystyle\rho_{yy}(q)=\frac{4\pi}{k_Fl}\frac{\hbar}{e^2},& 
{\rm for}~~~q\ll \frac{2}{l},
\end{eqnarray}
where $l$ is the CF transport mean free path at $\nu=\frac{1}{2}$. This
linear dependence of the conductivity $\sigma_{xx}(q)$, for $q\gg 
\frac{2}{l}$ is precisely what is needed to explain the SAW anomaly at $
\nu=\frac{1}{2}$ in the experiment of Willett et al \cite{will}. However, 
Halperin et al \cite{HLR} emphasized that there is no adjustable 
parameters in (\ref{clear}) but the theoretical values obtained 
from (\ref{clear}) is approximately a factor of 2 
smaller than the experimental values obtained by Willett et al. The main task
of present work is to try to resolve this systematic discrepancy between 
theory and experiment. 

Note that we use $K^0$ replaces $\tilde K$ in order to obtain (\ref{clear}).
It seems difficult to calculate the contribution from  diagrams other than 
the free CF response function in the Coulomb gauge.  
Now, we work in the temporal gauge and anticipate to find a comparable
contribution other than the free CF response. We consider a bubble diagram 
of the self-interaction among the gauge fluctuations (Fig. 1),
\begin{eqnarray}
K^a_{22}(\vec{q},\omega)&=&\int \frac{d^2q'}{(2\pi)^2}\frac{d\omega}{2\pi i}
(f_{2ab}f_{2a'b'}\nonumber\\
&\times&D_{aa'}(\vec{q}'+\vec{q},\omega'+\omega)D_{b'b}(\vec{q},\omega').
\end{eqnarray}
Associating to the long wavelength limit and $\omega\ll v^*_F q\to 0$, we
use the RPA propagator (\ref{prop}) for the gauge fluctuations. We then have
\begin{eqnarray}
K^a_{22}(q,\omega)&=& -\frac{1}{(8\pi m_b)^2}\int \frac{d^2q'}{(2\pi)^2}
\frac{d\omega}{2\pi i}\nonumber\\
&\times&(-q^2D_{11}D_{11}+q^{'2}_2D_{11}D_{22}- q^{'2}_2D_{22}D_{11}
)\nonumber \\
&\approx& \frac{i}{128\pi m^2_b m^*k_F}\frac{\gamma^2}{\chi^2}\int
q^{'2}d q'\nonumber\\
&=& \frac{ik_F^4 \varepsilon^2}{24\pi m_b^2 m^* e^4}.
\end{eqnarray}
In the last equality, we have truncated the wavevector of the gauge
fluctuations in $q<k_F$ \cite{SM}. Now, $\tilde K$ in (\ref{rhoyy}) 
can be approximated by $K^0+K^a$. Hence, the inverse
of the transverse resistivity of CF, in the clear range, is given by
\begin{eqnarray}
\frac{1}{\rho_{yy}(q)}&=&-e^2\lim_{\omega=v_sq\to 0} \frac{1}{\omega}
{\rm Im}(K^0_{22}(q,\omega) +K^a_{22}(q,\omega))\nonumber\\
&\approx& \frac{k_F}{4\pi q}(2-C)\frac{e^2}{\hbar}, \label{nrhoyy}
\end{eqnarray}
where we have restored the unit and the constant $C$ is defined as
\begin{equation}
C=\frac{v^*_F/v_s}{6(m_b/m_{coul})^2}= \frac{k_F\hbar/m^*v_s}{6(m_b/m_{coul})^2}
\label{C}
\end{equation}
with $m_{coul}=\frac{\varepsilon k_F\hbar^2}{e^2}$ being a mass scale induced by
the Coulomb interaction. Eq. (\ref{nrhoyy}) is our main result presented
here. Therefore, we can have an adjustable parameter $C$ in the 
$q$-dependent conductivity $\sigma_{xx}(q)$. In the following, we take two  
methods to compare our result with experiments. 

First, we take the effective mass in its theoretical value given by
(\ref{em}). For GaAs/AlGaAs heterostructres, we take the dielectric
constant $\varepsilon=12.6$ and the electron band mass $m_b\approx 0.07 m_e$. 
We refer to several sets of experimental data
by Willett et al as follows. (A) The 2DEG density $n_e=6.6\times 10^{10} 
{\rm cm}^{-2}$, the frequency of SAW, $f=2.4 {\rm GHz}$ and the corresponding wavelength 
$\lambda=1.2 \mu {\rm m} $ \cite{will}; (B) $n_e=6\times 10^{10} 
{\rm cm}^{-2}$, 
$f=1.5 {\rm GHz}$ and $\lambda=2.0 \mu {\rm m}$ \cite{will2}; 
(C) $n_e=7\times 10^{10} {\rm cm}^{-2}$, $f=0.36 {\rm GHz}$ and 
$\lambda=7.8\mu {\rm m}$ \cite{will5};
(D) $n_e=1.0\times 10^{11} {\rm cm}^{-2}$, $f=1.2 {\rm GHz}$ 
and $\lambda=8\mu {\rm m}$
\cite{will3}; and (E) $n_e=1.6\times 10^{11} {\rm cm}^{-2}$, 
$f=10.7 {\rm GHz}$
and $\lambda=0.27 \mu {\rm m}$ \cite{will4}. Corresponding to these experimental 
parameters, we have the constants $C$ as 
\begin{eqnarray}
C_A\approx 1.22, & C_B\approx 1.08, & C_C\approx 1.35, \nonumber\\
C_D\approx 0.56, & C_E\approx 3.00. &  ~~ \label{CN}
\end{eqnarray}
The first three in (\ref{CN}) are fairly good in comparing to $C\approx 1$.
It is reasonable that those constants are bigger than one because the 
CF effective mass is always enhanced by the gauge fluctuations in the real 
cases. The last one is obviously too big because it leads to a negative
$\sigma_{xx}(q)$. It implies that an effective mass much bigger than the
theoretical one (\ref{em}) is required (see below).
The forth constant seems to be smaller. 
This may be caused by either a smaller effective mass or 
a relative big ratio of $v_s/v^*_F$ such that 
the high order contribution and high loop diagrams have to be taken into 
account. We also would like to comment more on the exact quantitative 
comparison between theory and experiment. This kind of comparison, 
actually, is very difficult to pursue. The theoretical model we have 
employed is based on an ideal, infinitely thin 2DEG system in the absence 
of disorder and Landau level mixing whereas the 2DEG in GaAs/AlGaAs 
heterostructures has a non-zero thickness, finite carrier scattering 
time rate and is exposed to a finite magnetic field.

Second, we use a phenomenological CF effective mass $m^*_{ph}$ to replace
$m^*$ and write $m^*_{ph}$ as
\begin{equation}
\frac{1}{m^*_{ph}}=A_{ph}\frac{e^2}{\varepsilon k_F}.
\end{equation}
To resolve the discrepancy between theory and experiment, it has to set 
$C\approx 1$. 
Plugging the experimental data (A)-(E) into (\ref{C})
and requiring $C\approx 1$, we have 
\begin{eqnarray}
A_{ph,A}\approx \frac{1}{7}, & \displaystyle A_{ph,B}\approx \frac{2}{13},
& \displaystyle A_{ph,C}\approx \frac{1}{8},\nonumber\\
A_{ph,D}\approx \frac{5}{17}, &  \displaystyle  A_{ph,E}\approx \frac{1}{18},& 
\end{eqnarray}
where the capital letters in the suffices are respect with to the 
experimental data quoted above. We see that the first three phenomenological
data are fairly consistent with our theoretical value $A=\frac{1}{6}$ in 
a rate $<25\%$. It is understandable that the phenomenological effective 
mass is bigger than the theoretical value in the Hartree-Fock approximation 
since the gauge fluctuations always enhance the CF effective mass.
The forth one approximates to 0.3 which is in agreement with the value of the 
effective mass estimated by Halperin et al \cite{HLR}. And the last one 
leads to a much larger effective mass. So, we suggest experimentalists to check the   
the CF effective mass given by $C\approx 1$ in (\ref{C}) with other measurements 
to the effective mass for the same sample. The result $C\approx 1$ also 
suggests that there is the same $n_e$-dependence for both $m^*$ and $v_s$ 
in a wide range of the experimental data if $\nu=\frac{1}{2}$ is fixed. 
This requires further examinations both in experiment and in theory.

In conclusion, we have proposed a systematic method to resolve the discrepancy
between theory and experiment in the longitudinal conductivity dependent
on the wavevector of the SAW propagating above the 2DEG. The key point is
to take the correction to the response function from the 
self-interaction among the gauge fluctuations into account. This correction gives
an adjustable parameter to the conductivity. The phenomenological effective  
mass of CF can be estimated by using the data in the established experiments
and is found fairly consistent with our theoretical calculation in the 
Hartree-Fock approximation for some sets of the data. On the other hand, 
there was a discrepancy between the theoretical and  phenomenological values 
for one set of data, which may be caused by the correction to the effective 
mass from the long wavelength gauge fluctuations. Actually, we have suggested
a method to measure the CF effective mass and then the model presented here
could be checked by other measurements for the effective mass.

This work was supported in part by the NSF of China.


\centerline{\bf FIGURE CAPTIONS}

Fig. 1: One-loop diagram of the self-interaction of the gauge fluctuations.

Fig. 2: The self-interaction vertex of the gauge fluctuations.

\end{document}